**Superconductivity in the solid phases of Bi. Is Bi-IV a superconductor?**

Ariel A. Valladares[1] *, Isaías Rodriguez[2], David Hinojosa-Romero[1], Alexander Valladares[2], and Renela M. Valladares[2]

[1]Instituto de Investigaciones en Materiales, Universidad Nacional Autónoma de México, Apartado Postal 70-360, Ciudad Universitaria, CDMX, 04510, México.

[2]Facultad de Ciencias, Universidad Nacional Autónoma de México, Apartado Postal 70-542, Ciudad Universitaria, CDMX, 04510, México

*Corresponding author; e-mail: valladar@unam.mx

**The first successful theory of superconductivity was the one proposed by Bardeen, Cooper and Schrieffer in 1957. This breakthrough fostered a remarkable growth of the field that propitiated progress and questionings, generating alternative theories to explain specific phenomena. For example, it has been argued that Bismuth, being a semimetal with a low number of carriers, does not comply with the basic hypotheses underlying BCS and therefore a different approach should be considered. Nevertheless, in 2016 based on BCS we put forth a prediction that Bi at ambient pressure becomes a superconductor at 1.3 mK [1]. A year later an experimental group corroborated that in fact Bi is a superconductor with a transition temperature of 0.53 mK [2], a result that eluded previous work. So, since Bi is superconductive in almost all the different structures and phases, the question is why Bi-IV has been elusive and has not been found yet to superconduct? Here we present a study of the electronic and vibrational properties of Bi-IV and infer its possible superconductivity using a BCS approach. We predict that if the Bi-IV phase structure were cooled down to liquid helium temperatures it would also superconduct at a $T_c$ of 4.25 K.**

Bardeen, Cooper and Schrieffer (BCS) explained superconductivity by invoking two important concepts: The phonon-mediated electron Cooper pairing that occurs due to the vibrations in the material, giving rise to the transition to the superconducting state, and the coherent motion of the paired electrons that gives them the inertia to sustain electrical currents for a long time without dissipation. Simple but revolutionary. Several variations of these ideas have appeared in the course of time and even different concepts that pretend to substitute the original ones. Since vibrations are invoked to be the main factor leading to a bound electron pair, some manifestation of such interaction should appear in the phenomenon, and it does: the isotope effect. The Meissner effect is also duly accounted for and then the two main aspects of superconductivity are borne out by the BCS theory. Superconducting-like phenomena have been invoked in other realms of physics like nuclear and elementary particles where the pairing mechanism should be adequately chosen. It has also been ventured that in principle all materials may become superconductors if cooled down to low enough temperatures. We here show that invoking the corresponding electron and vibrational densities of states we can predict superconductivity, provided the Cooper attraction sets in. This elemental approach, if proven correct, would indicate that superconductivity in bismuth can be understood in a simple manner without invoking eccentric mechanisms.

In a very recent work [1] we computationally generated an amorphous structure of bismuth (*a*-Bi), characterized its topology, showed that it agreed remarkably well with experiment and then proceeded to calculate its electronic, $N(E)$, and vibrational, $F(\omega)$, densities of states to study their effect on the superconducting properties of this amorphous Bi phase. By comparing these results with the corresponding ones for the crystalline (Wyckoff) structure at atmospheric pressure we predicted that the crystalline material should become a superconductor at a temperature $T_c \leq$ 1.3 mK [1]. A year later an experimental group reported that, in fact, the Wyckoff phase is superconductive with a transition temperature of 0.53 mK [2], in agreement with our prediction. Encouraged by this success we decided to undertake a systematic study of the superconductivity of the solid phases of Bi under pressure and, in this paper, we put forth another prediction: the solid phase of bismuth known as Bi-IV, hitherto considered non-superconducting, should become a superconductor with a transition temperature close to the boiling point of liquid helium: 4.2 K. Since experimentalists are very ingenious, it may be possible to corroborate our prediction by applying a pressure of about 4 GPa on the Wyckoff structure and then rapidly cooling it down to below 4.2 K where the material should become a superconductor, always maintaining the under-pressure structure.

Solid Bi phases under pressure have been studied for decades. The pioneering work by Bridgman [3, 4] established the existence of several phases [5, 6] that have evolved into 5 presently accepted phases somewhat different from the original classification, Fig. 1. All these phases are crystalline (one claimed to be incommensurate, Bi-III) and it is now taken for granted that these 5 phases extend to low temperatures, except for Bi-IV that exists in a very well-defined area in the P-T plane, 2.5 GPa < P(Bi-IV) < 5.5 GPa, and in the neighborhood of 500 K. Bundy [6], in 1958, identified this phase and it was first assumed to be cubic.

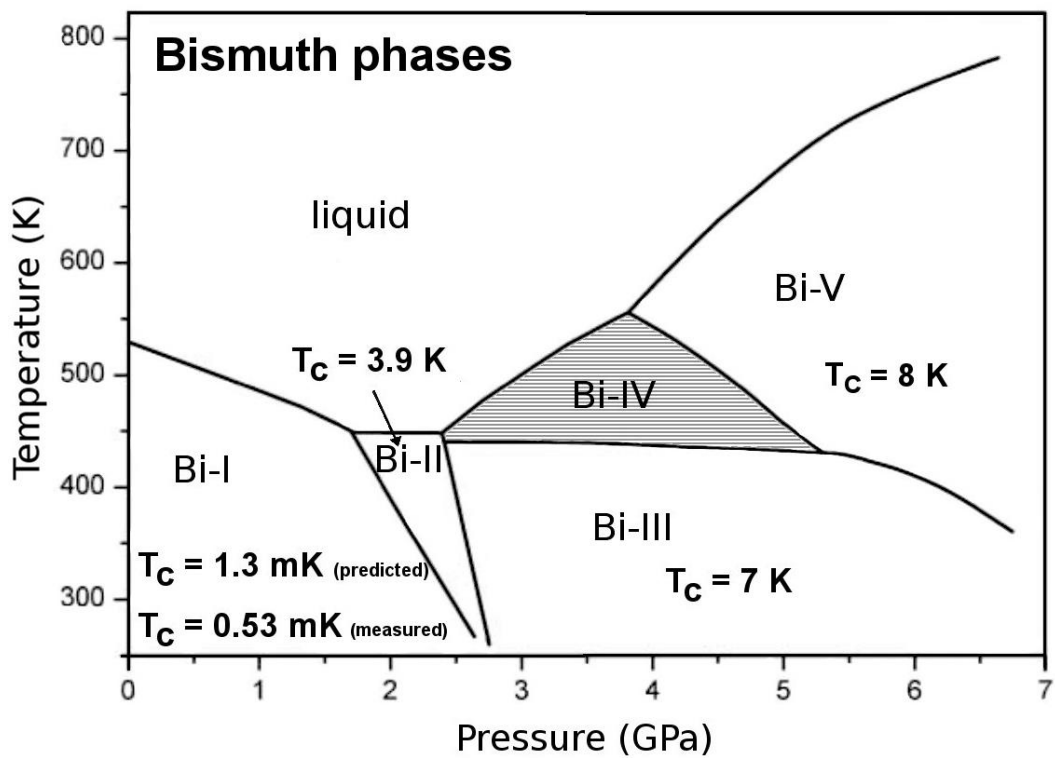

**Figure 1**. **Accepted present-day phases for bismuth as a function of pressure and temperature.** The known superconducting transition temperatures are indicated, [7]. For Bi-I we include the value we predicted [1] and the one measured [2].

By now, all of them except Bi-IV have been found to superconduct. Bi-I has a superconducting transition temperature of ~1 mK [1, 2]; Bi-II has a $T_c$ of 3.9 K; for Bi-III the transition temperature is 7 K and for Bi-V it is 8 K [7]. Under these circumstances Bi-IV seems to be out of place since no superconducting transition temperature has been reported for this structure; however, it may be possible that superconductivity in Bi-IV has not been found since this phase does not exist at low temperatures. In Fig. 1 a present-day classification of Bi phases with the enigmatic Bi-IV in the center of the plot [8, 9] is shown. In this figure an increasing tendency for $T_c$ seems to exist as the pressure is increased, at least up to 7 GPa, so one would expect Bi-IV to be a superconductor with a transition somewhere between 4 and 8 K. Would this surmise be true and if so, how can we calculate its transition temperature?

The fact that Bi-IV exists at high temperatures does not imply that a superconducting transition would occur at these temperatures, since this phenomenon would have been observed by now; so, room temperature superconductivity is ruled out. Then, what makes other phases superconductive while Bi-IV does not seem to be? We claim that Bi-IV would be superconductive if it were possible to quickly quench the structure to low temperatures avoiding structural changes. Herein we present results of what the transition temperature would be using a BCS approach and calculating from first principles the vibrational and electronic densities of states.

But, what is bismuth? A puzzling material; a versatile substance; it is the heaviest element of group 15, the highest atomic-number semimetal. At ambient pressure and temperature Bi is a crystalline solid, frustrated since it would like to be cubic but ends up being a rhombohedral (layered-like) structure, a semimetal for which the conducting properties are limited. At low temperatures *a*-Bi exists and becomes a conductor, but even better, it becomes a superconductor; it would seem that lifting all the symmetry restrictions of the Wyckoff crystal [10] frees the electrons and lets them Cooper-pair.

Bismuth under pressure changes structures and those associated to the 5 crystalline, solid phases have been determined experimentally, including the so-called incommensurate one of Bi-III. For our purposes, we use the recent experimental results for Bi-IV by Chaimayo *et al.* [9] that show that the structure, at 3.2 GPa and 465 K, is orthorhombic (pseudo-tetragonal) with the *oS*16 crystal structure and with the following lattice parameters: ***a*** = 11.191*(5)* Å, ***b*** = 6.622*(1)* Å and ***c*** = 6.608*(1)* Å. The space-group was confirmed as *Cmce* and is isostructural with Cs-V and Si-VI. Figure 2 represents this 16-atom structure with bilayers (white spheres) intercalated by gray and black monolayers. We constructed supercells by multiplying this 16-atom cell accordingly.

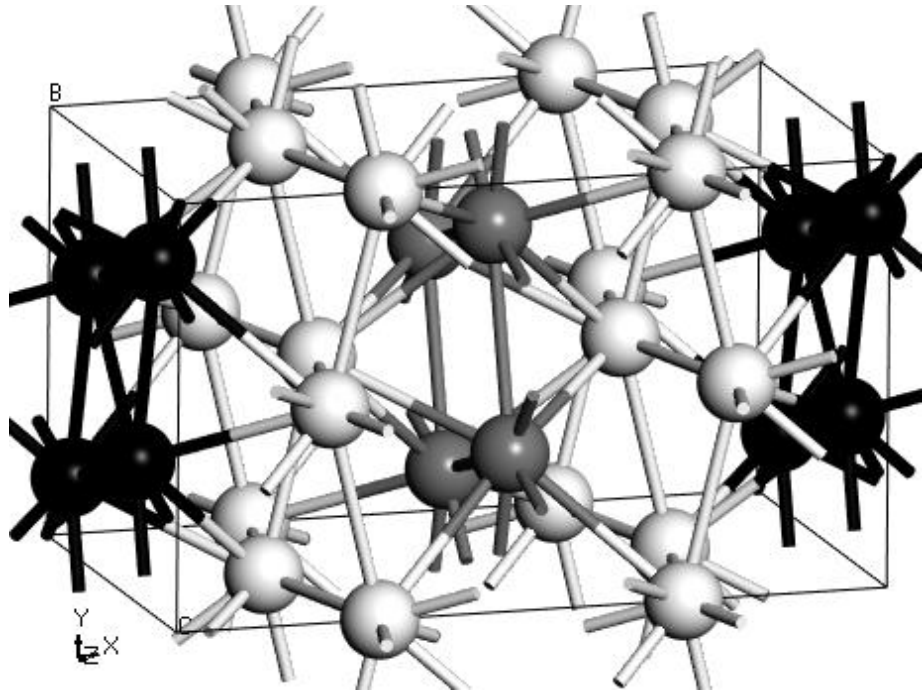

**Figure 2**. **Crystalline 16-atom cell of Bi-IV.** The supercell constructed to calculate $N(E)$ has 256 atoms and is obtained by multiplying 4x2x2 times the one depicted. The supercell used to calculate $F(\omega)$ has 128 atoms, a result of a 2x2x2 multiplication.

Next, we proceed to calculate and analyze the $N(E)$ and $F(\omega)$ for the Wyckoff and Bi-IV phases looking for a justification to validate our surmise that Bi-IV may become a superconductor. We shall estimate its superconducting transition temperature *a la* Mata *et al*. [1].

$N(E)$ and $F(\omega)$, the electronic and vibrational densities of states respectively, were calculated using the DMol3 code which is part of the Dassault Systèmes BIOVIA Materials Studio suite [11]. A single-point energy calculation was performed first using a double-numeric basis set and a finer mesh within the LDA-VWN approximation; an unrestricted spin-polarized calculation for the energy was carried out. Since bismuth is a heavy element with many electrons, the density-functional semi-core pseudo-potential (DSPP) approximation was used [12]. This pseudo-potential has been investigated by Delley where an all electron calculation is compared to the DSPP; the rms errors are essentially the same for both methods, 7.7 vs 7.5 [12]. Scalar relativistic corrections are incorporated in these pseudopotentials, essential for heavy atoms like Bi. Since DSPPs have been designed to generate accurate DMol3 calculations, it is expected that their use represents a good approximation; considerations of symmetry were left out for both structures. The parameters used in the calculations were the same for the respective $N(E)$ and $F(\omega)$, so meaningful comparisons can be made. For example, an energy convergence of $10^{-6}$ eV was used throughout, the real space cutoff was set to 6.0 Å, and the integration grid was set to fine, so the calculations were carried out using a Monkhorst-Pack mesh of 2x3x3 in k-space. For the vibrational calculations, the finite-displacement approach was employed with a step size of 0.005 Å to calculate the Hessian using a finite-difference evaluation.

We now base our discussion on the BCS expression for the transition temperature:

$$T_c = 1.13\, \theta_D\, [\exp(-1/N(E_F)V)], \qquad (1)$$

where $\theta_D$ is the Debye temperature and represents the role played by the vibrational density of states, vDoS, typified by $F(\omega)$. $N(E_F)$ is the electron density of states, eDoS, at the Fermi level $E_F$, and $V$ is the Cooper pairing potential that binds pairs of electrons [13]. The dependence of $T_c$ on the parameters is represented in Fig. 3 for a specific value of the pairing potential where the strong dependence of the transition temperature on the factor $N(E_F)$ for a given value of $V$ can be observed [14].

This indicates that under certain circumstances $N(E_F)$ can play a more important role than the vibrations, especially when different phases of the same material are compared since in this case it can be assumed that the strength of the pairing potential would not be altered much by the phase changes. Although the Debye temperatures may not change drastically, the electronic properties could change radically going from being a semimetal to becoming a conductor in the case of Bi.

The calculations were carried out as follows. For the Wyckoff structure, we used a 6-atom crystalline cell [10] that we multiplied 5x4x2 times to obtain a supercell with 240 atoms for $N(E)^W$ and 3x3x2 times to obtain a supercell with 108 atoms to calculate $F(\omega)^W$. For the Bi-IV structure we considered the experimental 16-atom cell by Chaimayo *et al.* [9] mentioned above, and depicted in Fig. 2, and the supercell used was a (4x2x2) x 16 = 256-atom one for the calculation of $N(E)^{IV}$ and for $F(\omega)^{IV}$ a 128-atom supercell (2x2x2 times the Chaimayo cell) was used.

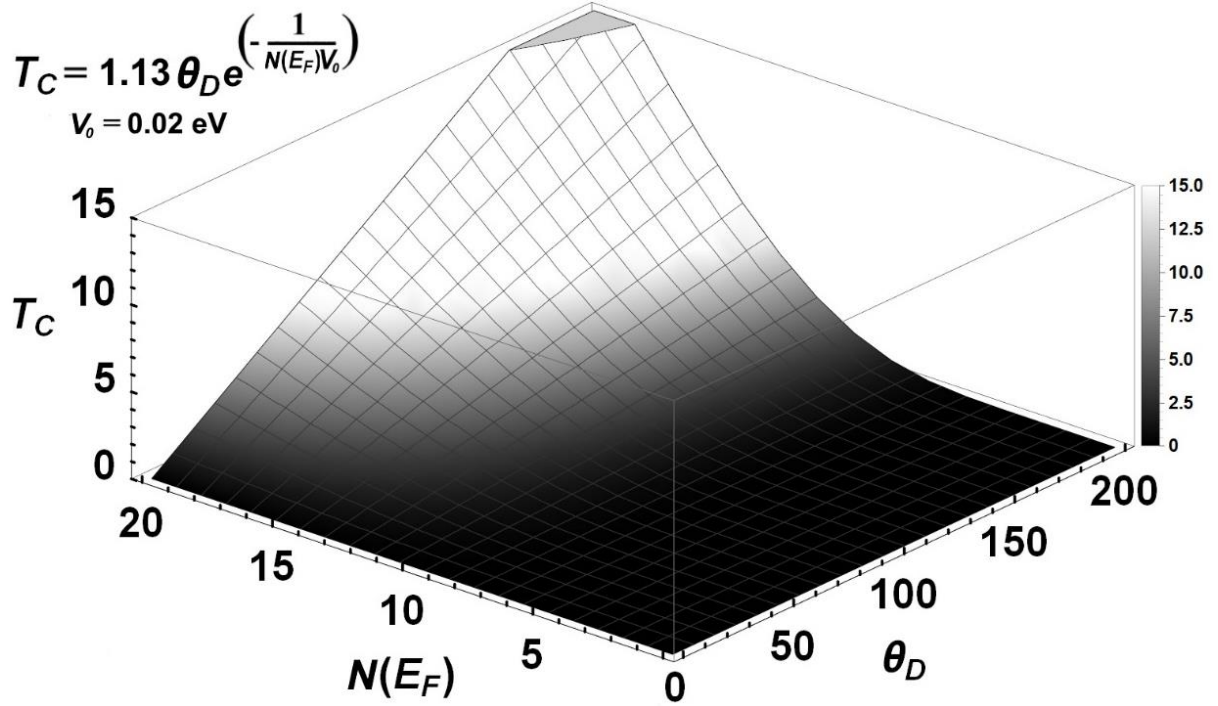

**Figure 3. Dependence of the superconducting transition temperature $T_c$ on the electronic $N(E_F)$ and vibrational $\theta_D$ parameters.** The pairing potential $V_o = 0.02$ eV was chosen arbitrarily.

For the eDoS the DMol3 analysis tools included in the Materials Studio suite were used, set to eV, and also an integration method with a smearing width of 0.2 eV. The number of points per eV was 100. The results for the densities of states are given per atom in Fig. 4.

For the vDoS the results were analyzed with the OriginPro software, the normal modes calculated were imported in THz. To obtain the vDoS a frequency count with a 0.11 THz bin width was used, and the resulting bins were smoothed with a two-point FFT filter. The three translational modes around 0 THz were removed. The results for the densities of states are given per atom in Fig. 5.

Previously we performed some estimations of the critical temperature of the Wyckoff phase by calculating from first principles the eDoS and the vDoS and correlating them to the corresponding values of the amorphous phase [1]. We use the same successful approach for the Wyckoff and the Bismuth IV phases, using the results depicted in Figs. 4 and 5, to investigate the possible existence of superconductivity in Bi-IV.

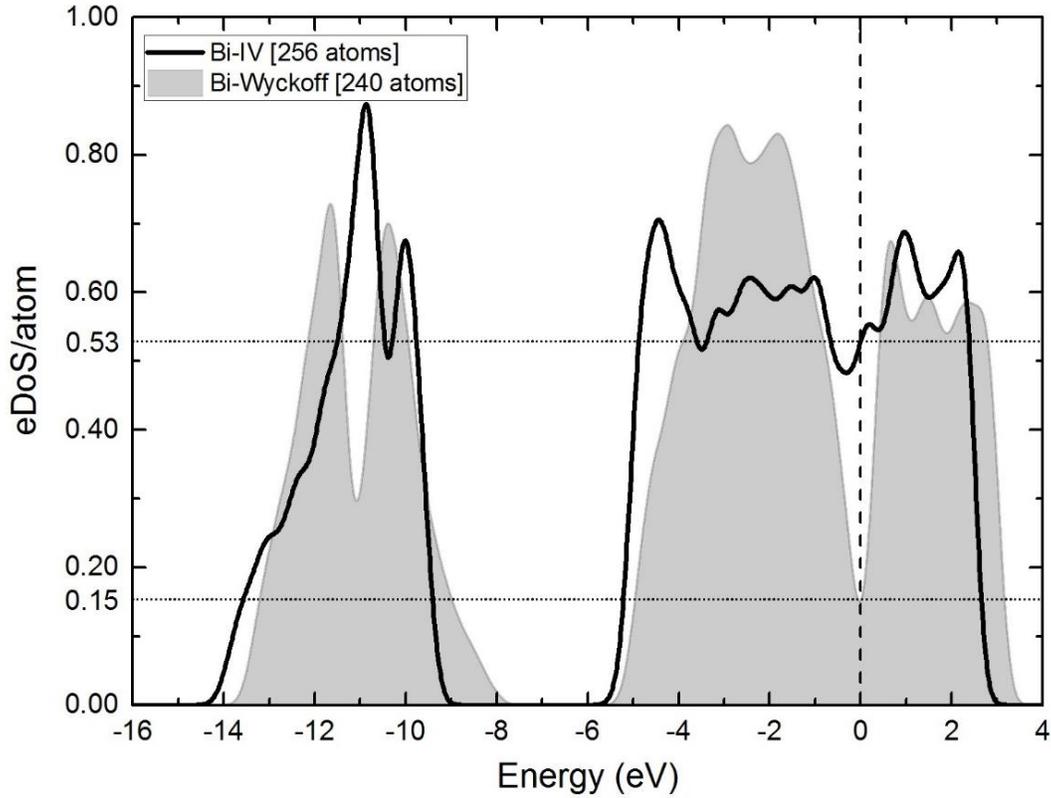

**Figure 4**. **Comparison of the electron densities of states for the Wyckoff (gray plot) and Bi-IV (black line) phases.** The vertical axis is $N(E)$/atom so that we can compare supercells of different sizes. $E_F$, the Fermi energy, is the reference value for the energy.

Figures 4 and 5 present comparisons between the results for Bi-IV and for the Wyckoff structure for $N(E)$ and $F(\omega)$, respectively. We tried to use supercells with a similar number of atoms for each calculation, but since the computational resources to obtain $F(\omega)$ are very demanding we ran some tests to find an equilibrium between the accuracy of the results and the size of the supercells used. That is why the number of atoms in the cells utilized to calculate $F(\omega)$ is smaller than those used to calculate $N(E)$. Both electronic and vibrational properties are essential and are represented by the results depicted in Figs. 4 and 5.

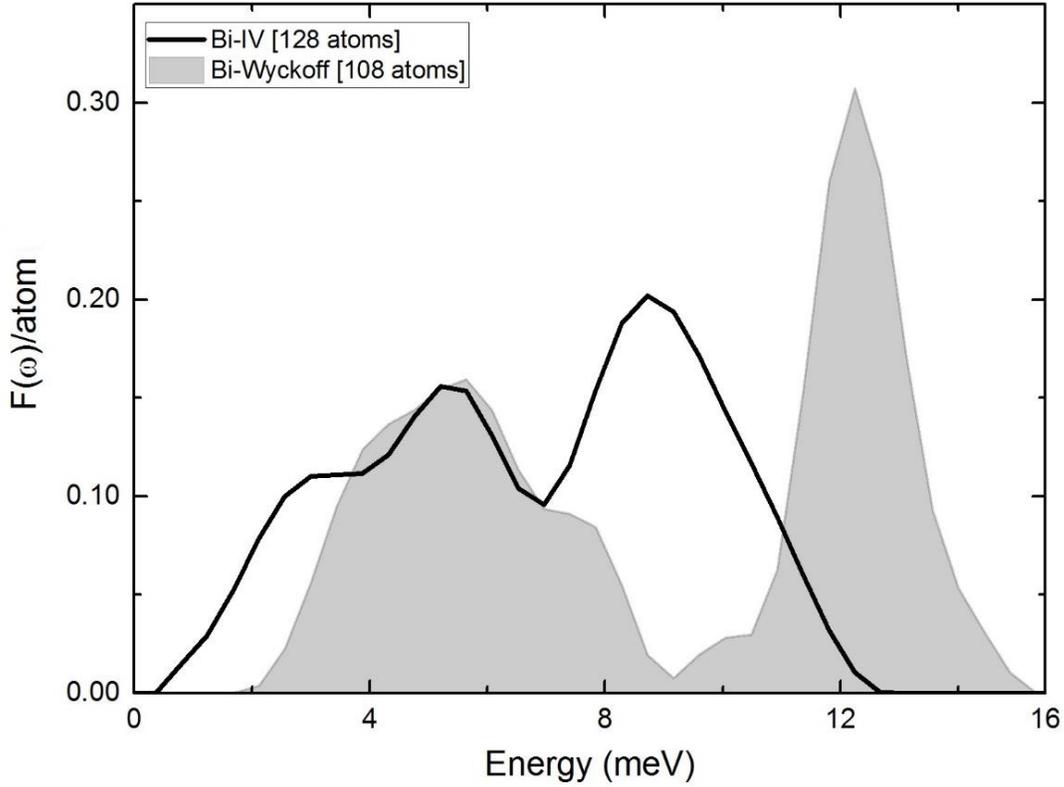

**Figure 5. Comparison of the vibrational densities of states for the Wyckoff $F(\omega)^W$ (gray plot) structure and for Bi-IV $F(\omega)^{IV}$ (black line) phase.** Since the supercells have a different number of atoms, to compare them we plot the $F(\omega)$ per atom.

From Figure 4 the values $N(E)^{IV}$ and $N(E)^W$ for the Bi-IV and the Wyckoff phases respectively, can be analyzed. The $N(E)^{IV}$ has a high number of electron states, 0.53, at the Fermi level indicating a metallic character. $N(E)^W$, as expected, is lower in comparison since this phase is semimetallic, 0.15. The ratio of the densities of electron states at the Fermi level for these two phases is 3.53.

To incorporate the Debye temperatures as required by Eqn. (1) we need to calculate them from the vibrational spectra presented in Figure 5. For this we use an expression due to Grimvall [15],

$$\omega_D = \exp\left[ 1/3 + \frac{\int_0^{\omega_{max}} \ln(\omega) F(\omega) \, d\omega}{\int_0^{\omega_{max}} F(\omega) \, d\omega} \right], \quad (2)$$

that we utilized in Ref. 1 with good results. $F(\omega)$ is the vDoS of the supercell under consideration, Figure 5, and $\omega_{max}$ is the maximum frequency of the vibrational spectrum. Using Eqn. 2, the calculations for $\theta_D = \hbar\omega_D / k_B$ give: $\theta_D^{IV} = 102.1$ K and $\theta_D^W = 134.2$ K, with $k_B$ the Boltzmann constant; the ratio being 0.76. An observation is pertinent, the vibrational spectrum for Bi-IV is more localized than the spectrum of the Wyckoff phase, although one would expect it to be more extended since the application of pressure would bring the atoms together, but evidently the change in the crystalline structure supersedes this fact. The density on the other hand does increase, 9.8 *vs* 11.33 g/cm$^3$.

We should mention that to obtain these results we removed the translational modes ($\omega \approx 0$) that are more preponderant the smaller the number of atoms in a supercell. That is why our present numbers are somewhat larger than those reported in Ref. 1 for the Wyckoff phase, 129 K. The experimental values for $\theta_D$ reported by DeSorbo [16] for crystalline bismuth at ambient pressure varies from 140 K at high temperatures to 120 K at low temperatures. Our calculations indicate that $\theta_D$ for the crystal at ambient pressure lies between the experimental values so we trust the results of 134.2 K for Wyckoff and 102.1 K for Bi-IV.

Now we may compare the possible $T_c$ of the Bi-IV phase with either the $T_c$ of the amorphous or with the $T_c^W$ of the Wyckoff phase that was recently discovered to be superconductive. To be consistent, we do it with the superconductivity we predicted for the Wyckoff phase in the BCS approach. Suppose that superconductivity is possible for Bi-IV, with a superconducting transition temperature $T_c^{IV}$, and assume that the Cooper pair potential $V$ is essentially the same for these two phases of the same material. Then the transition temperatures are:

$$T_c^{IV} = 1.13\, \theta_D^{IV}\, exp\,(-1/\, N(E_F)^{IV}\, V)$$

$$T_c^W = 1.13\, \theta_D^W\, exp\,(-1/\, N(E_F)^W\, V).$$

If in general, we assume that

$$N(E_F)^{IV} = \alpha\, N(E_F)^W \qquad \text{and} \qquad \theta_D^{IV} = \beta\, \theta_D^W,$$

we can rewrite the expression for $T_c^{IV}$ as:

$$T_c^{IV} = \{T_c^W\}^{1/\alpha}\, \beta\, \{1.13\, \theta_D^W\}^{(\alpha-1)/\alpha},$$

and substituting the values for $α = 3.53$ and $β = 0.76$ for our case, we obtain

$$T_c^{IV} = 4.25 \text{ K}.$$

In conclusion, predicting the existence of superconducting phases of Bismuth is a risky endeavor since calculating the pairing potential is a challenge. Computational simulations may help to accomplish this but meanwhile we can obtain reasonable approximations to the superconducting transition temperatures if one deals with the same material in different phases, case in point.

Also, computational simulations have evolved favorably to allow one to obtain reliable results for the electronic and vibrational densities of states to be able to compare them meaningfully and infer conclusions. In this work, it is evident that for electrons, Fig. 4, going from the Wyckoff phase, clearly a semimetallic phase, to the Bi-IV phase, clearly a conducting phase, implies a very definite change of behaviour in the electronic transport properties. As far as the vibrational properties are concerned, one can observe changes that indicate that for Bi-IV the existence of low frequency modes is notable and could re-assert the often invoked characteristic that the soft phonon modes are important in the manifestation of superconducting properties, Fig. 5. The Wyckoff phase even if superconducting, and accepting the same pairing potential, must work against the lower $N(E)$ and the absence of soft phonons to display superconductivity at more accessible temperatures. Bi-IV does not have to work much since with the same pairing potential it has more electrons at $N(E_F)$ to pair and more soft phonons to do the pairing. One drawback is the extension of the $F(ω)$ which is smaller for Bi-IV than for Wyckoff, Fig. 5. Also, interesting to note is that the gap in the $F(ω)$ that exists for Wyckoff disappears for Bi-IV This may be because there are monolayers intercalated between the by-layers in the crystalline structure of Bi-IV, Fig.2.

Certainly, the electronic and vibrational properties are conducive to propitiating superconductivity; what is needed to detonate it is that the pairing potential manifests itself and for that the temperature has to be lowered until superconductivity may appear at liquid helium temperatures, maintaining the Bi-IV structure, as we predict in this work.

Perhaps it may sound daring to put forth the hypothesis that BI-IV may become a superconductor, especially when our number of successes is limited to 1, the Wyckoff phase. However, we have been following this line of thought and have found that the superconducting transition temperatures of the other solid Bi phases under pressure can be predicted and understood in this manner and we shall present these results in a future publication [17].

**Acknowledgments**

I. R. and D.H.R. acknowledge CONACyT for supporting their graduate studies. A.A.V., A.V. and R.M.V. thank DGAPA-UNAM for continued financial support to carry out research projects IN101798, IN100500, IN119908, IN112211, IN11014 and IN104617. M.T. Vázquez and O. Jiménez provided the information requested. Simulations were partially carried out in the Computing Center of DGTIC-UNAM.